# Article Commentary on "Microdosimetric and radiobiological effects of gold nanoparticles at therapeutic radiation energies" [T.M. Gray et al., IJRB 2023, 99(2), 308–317]

Hans Rabus[a]*, Miriam Schwarze[a], and Leo Thomas[a]

[a]Physikalisch-Technische Bundesanstalt (PTB), Berlin, Germany

*hans.rabus@ptb.de

In the recently published article by T.M. Gray et al. "Microdosimetric and radiobiological effects of gold nanoparticles at therapeutic radiation energies" (IJRB 2023, 99(2), 308–317) results of Monte Carlo simulations and radiobiological assays on the dosimetric effects of gold nanoparticles were presented. This commentary points out that the results of the two parts of the study are in contradiction and that the predicted magnitude of dose enhancement and its dependence on the shape of the nanoparticle appear implausible. Possible reasons for these observations are discussed.

**Introduction**

In their recently published work, Gray et al. (2023) performed radiobiological cell experiments to study the effects of the presence of gold nanoparticles (GNPs) during irradiation. In addition, Monte Carlo (MC) simulations were performed to determine the dose-enhancing effects of GNPs at the microscopic level. In these simulations, two different shapes (cubic, spherical) were considered for the same GNP volume. The spherical GNPs had a diameter of 30 nm, and the cubic GNPs had a side length of about 24 nm.

Irradiations were simulated for 6 MV and 18 MV linac radiation in a three-step procedure. In the first step, an experimental setup was simulated to obtain the photon fluence in the region of interest. This photon fluence was used in the second simulation step to irradiate the nanoparticles and obtain phase-space files of emitted electrons. In the third step, the energy imparted by these electrons to the water surrounding the nanoparticle was scored as a function of radial distance from the center of the nanoparticle.

Based on these simulations, dose enhancement factors (DEFs) were determined for a 1 μm³ volume of water containing the nanoparticle. The DEF values reported were about 6 and 8 for the sphere and the cube, respectively, with the 6 MV spectrum. For the 18 MV spectrum, the values were about 2.7 and 3.3 for the sphere and cube, respectively.

From the radiobiological assays, a radiosensitization enhancement factor (REF) was determined for given survival fractions (SFs) of cells. For the two SFs considered, 0.3 and 0.6, an REF of about 1.06 was determined for a mass fraction of gold nanoparticles of 0.10%. For a mass fraction of 0.15%, the REF was about 1.11.

*Observations on the paper*

It should be noted that the adjective "microdosimetric" is used in the work of Gray et al. (2023) to indicate that the absorbed dose was determined in a micrometric volume. This is not exactly what the term "microdosimetry", which is also a keyword of the paper, generally refers to, namely to the study of the stochasticity of ionizing radiation interaction at the microscopic scale (Rossi and Zaider 1996; Lindborg and Waker 2017).

Leaving this terminological issue aside, the nexus between the radiobiological part of the study and the Monte Carlo simulations is not immediately evident. The study of the difference between cubic and spherical nanoparticles was presumably motivated by the transmission electron microscopy images of the GNPs, which appear to exhibit a cross-sectional shape more like a square than a circle. However, it is not clear how the simulation results are related and relevant to the radiobiological assays. In fact, as will be explained in the next subsections, the results of the two parts of the study (MC simulations and radiobiological assays) appear to be in contradiction. Furthermore, the magnitudes of the local dose around the GNP and the dose enhancement and the influence of the shape of the nanoparticle on it are not plausible. Finally, the results for the radiobiological part of the study do not appear to be statistically significant.

*Contradictory results between MC simulations and radiobiological assays*

The results from the MC simulations and the radiobiological assays seem to be in contradiction for the following reason: A 30 nm diameter sphere has a volume of $1.4 \times 10^{-5}$ μm³. Therefore, the mass fraction of gold in 1 μm³ of water containing a GNP is about $2.7 \times 10^{-4}$ or 0.03 %, which is much lower than the mass fractions used in the radiobiological experiments of Gray et al. (2023).

This means that in a simulation relevant for the experiments, there should have been more than one GNP in a 1 μm³ volume, namely between four and six for 0.1 % and 0.15 % mass fraction, respectively. However, according to the simulations, already one GNP results in an enhancement of the dose in a 1 μm³ volume by a factor of about 3 for both shapes and the 18 MV irradiation. In the presence of four or six GNPs in the 1 μm³ volume instead of only one, the contribution from GNPs should be even further increased, and the resulting dose enhancement should be more than 3. Thus, instead of a dose of 3 Gy, a dose of more than 9 Gy would result in a volume with both 0.1 % and 0.15 % mass fraction of GNPs. The functional shape of the survival curve without GNPs shown in Fig. 7(a) of Gray et al. (2023) indicates that such high doses result in surviving fractions well below $10^{-2}$. Therefore, a much greater reduction in cell survival would be expected for the experiments with GNPs than is seen in Fig. 7(a) of Gray et al. (2023).

It should also be noted that the dependence of the dose enhancement factor on the primary radiation spectrum is contrary to the findings reported by Gray et al. (2021). There, measurements and simulations with an 18 MV linac spectrum were found to produce a larger DEF than a 6 MV spectrum. These finding were presumably the motivation for performing the cell survival studies at 18 MV in the work of Gray et al. (2023).

*Implausibility of the local dose per photon values*

A further implausibility is the magnitude of the local dose shown in Fig. 5 of Gray et al. (2023), which is between $1 \times 10^{-25}$ Gy and $3 \times 10^{-24}$ Gy in the range up to 500 nm from the center of the GNP. This suggests that the average dose in a 1 μm³ cube is in the order of a few $10^{-25}$ Gy. The mass of 1 µm³ water is $10^{-15}$ kg. Thus, the energy imparted is a few $10^{-40}$ J, or a few $10^{-21}$ eV. If the quantity shown in Fig. 5 of Gray et al. (2023) is normalized to the number of photons used in the second simulation ($4\times10^9$), the total energy scored in the 1 µm³ water volume would be in the order of $10^{-11}$ eV. This is obviously impossible since for each data point in Fig. 5 there must be at least one interaction (for the entire simulation), and the energy imparted by a single ionization is in the order of 10 eV.

*REF definition*

According to the Materials and Methods section in Gray et al. (2023), the REF is defined as follows:

> Radiosensitization enhancement factor (REF) is the ratio of the dose producing a given cell survival percentage in the presence of GNPs to the dose producing the same cell survival percentage in the absence of GNPs.

Since the presence of GNPs reduces the survival rate after irradiation, this definition implies that the REF values should be less than unity, whereas the reported values in their Table 1 are higher than unity. From the Results and Discussion sections, it is evident that Gray et al. (2023) used an REF definition analogous to those used by Chithrani et al. (2010), Kaur et al. (2013), and Cui et al. (2017). In these articles, the REF was defined as

$$REF_{x_g,S} = \frac{D_0(S)}{D_{x_g}(S)} \qquad (1)$$

where $D_0(S)$ and $D_{x_g}(S)$ are the dose values that produce a survival rate $S$ at mass fractions of gold of 0 and $x_g$, respectively. This quantity was defined by the International Commission on Radiation Units and Measurements (ICRU) as Dose Modification Ratio (DMR) (ICRU 1979). (It should be noted that Cui et al. (2017) used a slightly more intricate definition of the REF. However, when the same survival rates are considered for the absence of GNPs and their presence, the REF definition of Cui et al. (2017) is identical to the one given by Eq. (1).)

*Synergistic effects*

In Table 2 of Gray et al. (2023), results are presented for the reduction of survival by GNPs and radiation alone and for their simultaneous application. The latter gives larger effects than the combination of the first two. The data column for the case of "radiation only" contains values that vary with GNP concentration. This is implausible, since the "radiation only" data cannot be obtained from cells containing GNPs.

**Methodological concerns**

*Statistical significance of the change in cell survival in the presence of GNPs*

The p-values shown in Fig. 7(b) and (c) of Gray et al. (2023) appear implausible given the large (overlapping) error bars. These p-values and uncertainties of the observed REFs are not discussed in the paper.

In addition, it should be noted that the survival curves were determined by fitting the linear-quadratic (L-Q) two-parameter model to only two data points. Since the model value at dose zero is independent of the two model parameters, the inclusion of the data point at zero dose does not affect the fit results. Therefore, the best fit curve is effectively an interpolation of the two data points.

*Dependence of the dose enhancement on nanoparticle shape*

The reported DEF for the cubic nanoparticle is about 30% higher than the DEF for the spherical nanoparticle. This may be an artefact of the simulation geometry used in the second simulation step. In their Materials and Methods section, Gray et al. (2023) describe this second simulation step as follows:

> In a subsequent microscopic simulation, this energy spectrum was used for a set of six-point sources evenly distributed around a single GNP as seen in Figure 1(b). Each point source was placed 1 nm from the surface of the GNP.

Their Fig. 1(b) only shows the case of the spherical GNP. For the cubic nanoparticle, it seems reasonable to assume that the point sources were placed on the lines passing the GNP center and the centers of the faces. If this was the case, then the simulation setup may have been biased in favor of the cubic nanoparticle. This is illustrated in Fig. 1, which shows that the solid angles covered by beams emitted from the source at point P and hitting the GNP are different for the sphere and the cube.

Fig. 1(a) shows a cross section through the spherical GNP (circle) and the half-cone (dashed lines) of half-opening angle $\theta_{m,s}$, which delimits the solid angle within which all beams emitted from the point source at P intersect the GNP. This solid angle is given by Eq. (2).

$$\Omega_s = 2\pi(1 - \cos\theta_{m,s}) \qquad (2)$$

Unlike for the sphere, it is not possible to find an analytical expression for the solid angle subtended by the cube. However, it is possible to give upper and lower limits. This is illustrated by Fig. 1(b), which shows a view of the front of the cube as seen from the point source. All emitted beams incident on the plane of the front face within the long-dashed circle intersect the cube. Fig. 1(c) shows a cross section through the cube in the plane defined by points A, A', and P. Also shown is the semi-cone (dashed lines) of half-opening angle $\theta_{m,c}^*$, which bounds the solid angle subtended by the long-dashed circle in Fig. 1(b).

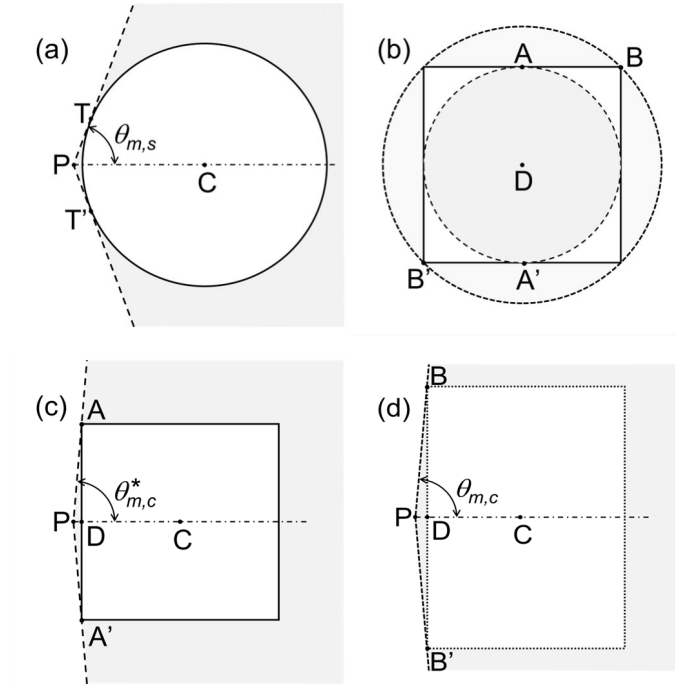

Fig. 1: Illustrations (to scale) of the irradiation geometry used in the simulations by Gray et al. (2023). (a) Cross-sectional view of the sphere and the solid angle (gray shaded area) covered by beams emitted from the source at point P that intersect the GNP. Point C is the GNP center. Points T and T' denote the tangential points of the dashed lines and the circle. (b) "Front view" of the irradiation of the cubic GNP. The long-dashed line indicates the boundary of the solid angle within which all beams emitted from the point source intersect the cube. The short-dashed line indicates the solid angle within which the beams intersect the cube for some azimuthal angles. Point D is the center of the cube face; points A and A' are the centers of two opposite edges of the cube face; B and B' are two opposite corners on that face. (c) and (d) Cross sections through the cube in the planes PAA' and PBB', respectively, and through the solid angles indicated by the long- and short-dashed lines in (b).

Beams intersecting the plane of the front face between the long-dashed circle and the short-dashed circle hit the cube only when they strike the white areas near the corners. Hence, the solid angle subtended by the short-dashed circle is an upper bound on the actual solid angle. Fig. 1(d) shows a cross section through the cube in the plane defined by points B, B', and P, and the cross section through the half-cone (short-dashed lines) with the half-opening angle $\theta_{m,c}$ corresponding to the short-dashed circle in Fig. 1(b). Therefore, the following relation holds for the solid angle for the cube:

$$2\pi(1 - \cos\theta^*_{m,c}) \leq \Omega_c \leq 2\pi(1 - \cos\theta_{m,c}). \tag{3}$$

In Eqs. (2) and (3), $\theta_{m,s}$, $\theta^*_{m,c}$, and $\theta_{m,c}$ are the half-opening angles of the conical boundaries of the solid angles shown in Fig. 1(a), (c), and (d), respectively. They are given by (cf. Fig. 2)

$$\theta_{m,s} = \sin^{-1}\frac{r}{r+d} \qquad \theta^*_{m,c} = \tan^{-1}\frac{s}{2d} \qquad \theta_{m,c} = \tan^{-1}\frac{s}{\sqrt{2}d} \tag{4}$$

where $r = 15$ nm is the radius of the sphere, $s \approx 24.2$ nm is the side of the cube, and $d = 1$ nm is the distance of the source from the GNP. The resulting solid angle for the sphere is 4.10 sr. The solid angle for the cube is between 5.77 sr and 5.92 sr, that is, more than 40% larger than for the sphere.

This means that photons emitted in the simulations from the point source have a 40% higher probability of intersecting the GNP when it is a cube than when it is a sphere. This may be the reason for the higher DEFs found for the cube. Since this bias is larger than the difference in DEF found between a spherical and a cube-shaped nanoparticle, it could even be that for unbiased results the DEF for the cube is smaller than for a sphere.

### *Purpose of this commentary*

The issues stated in the subsection "Observations on the paper" are based on simple plausibility arguments. The other points mentioned in the "Methodological concerns" subsection required somewhat more advanced treatment, such as geometric considerations. Beyond highlighting the issues and concerns, this commentary is meant to answer the following questions, which require more elaborate approaches than back-of-envelope calculations:

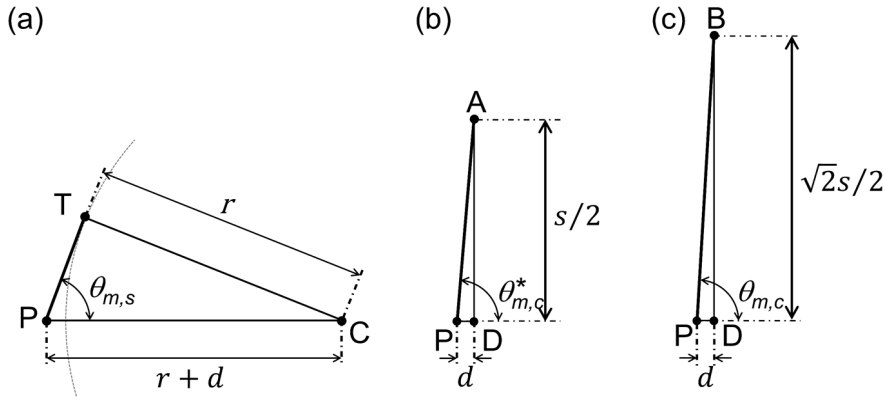

Fig. 2: Illustrations (to scale) of the triangles used for determining the solid angles in Fig. 1 (a), (c) and (d).

(1) What is a realistic value for the DEF in a 1 μm³ volume containing a GNP of the size considered in the study of Gray et al. (2023)?
(2) What is the order of magnitude of the local dose around such a GNP?
(3) What is the magnitude of the bias introduced by the point source geometry used by Gray et al. (2023) in their second simulation step?
(4) What are the uncertainties associated with the results of the radiobiological assays reported by Gray et al. (2023)?

The first two questions are addressed using information available in the literature to derive quantitative estimates. The third question is answered by evaluating the interaction probabilities of photons emitted from a point source for the two GNP shapes compared to the case of uniform isotropic irradiation. For the last question, the data presented by Gray et al. (2023) are reanalyzed including uncertainty propagation.

**Materials and Methods**

To investigate the issues mentioned in the introduction, an estimate of the maximum possible DEF for 6 MV linac irradiation was derived using a photon energy spectrum reported by McMahon et al. (2011). This photon spectrum was also used for estimating the expected magnitude of the local dose around the GNP. The chord length distribution and mean chord lengths for the irradiation geometry used in the second simulation step of Gray et al. (2023) were determined to assess a possible bias introduced by the simulated irradiation geometry. Finally, the data shown in Fig. 7 of Gray et al. (2023) were extracted and used to determine the REF values and their uncertainties.

*Estimate for the upper bound of the DEF in the simulations*

The dose to a uniform mixture of gold and water under secondary electron equilibrium was determined for the mass fraction of gold corresponding to a single GNP in a 1 μm³ volume of water. As was shown previously, this dose is an upper limit to the average dose to a volume of water around a nanoparticle under secondary electron equilibrium (Rabus et al. 2019). Therefore, the maximum possible DEF in a volume of water containing a nanoparticle is given by

$$DEF_{max} = (1 - x_g) + x_g \times \gamma_{g,w} \quad (5)$$

where

$$\gamma_{g,w} = \langle E \times (\mu_{en}/\rho)_g \rangle_{\Phi_E} / \langle E \times (\mu_{en}/\rho)_w \rangle_{\Phi_E}. \quad (6)$$

In Eqs. (5) and (6), $x_g$ is the mass fraction of gold, and $(\mu_{en}/\rho)_g$ and $(\mu_{en}/\rho)_w$ are the mass energy absorption coefficients of gold and water, respectively. $E$ denotes the photon energy, and the brackets $\langle \ \rangle_{\Phi_E}$ indicate a weighted average with respect to the spectral photon fluence $\Phi_E$. Under secondary particle equilibrium, this weighted average is the dose-to-fluence ratio. The estimate given by Eqs. (5) and (6) is an upper limit since some of the released energy is absorbed in the GNP. A procedure to correct the values for a homogenous mixture of gold and water for this absorption in the GNPs was developed by Koger and Kirkby (2016).

In essence, the line of reasoning leading to Eq. (5) is as follows: Assume secondary charged particle equilibrium and that the nanoparticles are arranged in a regular array of voxels as shown in Fig. 1(a) of Gray et al. (2023). Then the average energy imparted in a voxel 'A' by electrons produced by photons interacting in a voxel 'B' is the same as the average energy imparted in voxel 'B' by electrons produced in voxel 'A'. Therefore, the total energy imparted in a voxel can be estimated by the total energy transferred to electrons by photon interactions in that voxel.

The energy transferred by a photon of given energy is proportional to the mass energy transfer coefficient, which is approximately equal to the mass energy absorption coefficient. For a mixture of materials, the mass energy absorption coefficients have to be weighted by the mass fractions of the different components.

The photon energy spectra used in the microscopic simulation were not shown in the work of Gray et al. (2021, 2023). Therefore, Eq. (5) was evaluated using the photon spectrum reported by McMahon et al. (2011) for 6 MV linac radiation and 5 cm depth in water. The data from Fig. 1 in the paper of McMahon et al. (2011) were digitized using WebPlotDigitizer (https://apps.automeris.io/wpd/) and interpolated using GDL. The interpolated data were then fed into an Excel template developed in earlier work (Rabus et al. 2019). Visual Basic macro functions are implemented in the Excel workbook to calculate interpolated values of the mass energy absorption coefficients of gold and water for given photon energies. In the main worksheet, numerical integrals of the dose-to-fluence ratios are calculated for gold and water under secondary particle equilibrium. It should be noted that it was found in this procedure that the photon fluence shown in Fig. 1 of McMahon et al. (2011) is the fluence per eV and Gy cm² and not per keV and Gy cm² as stated in the figure caption.

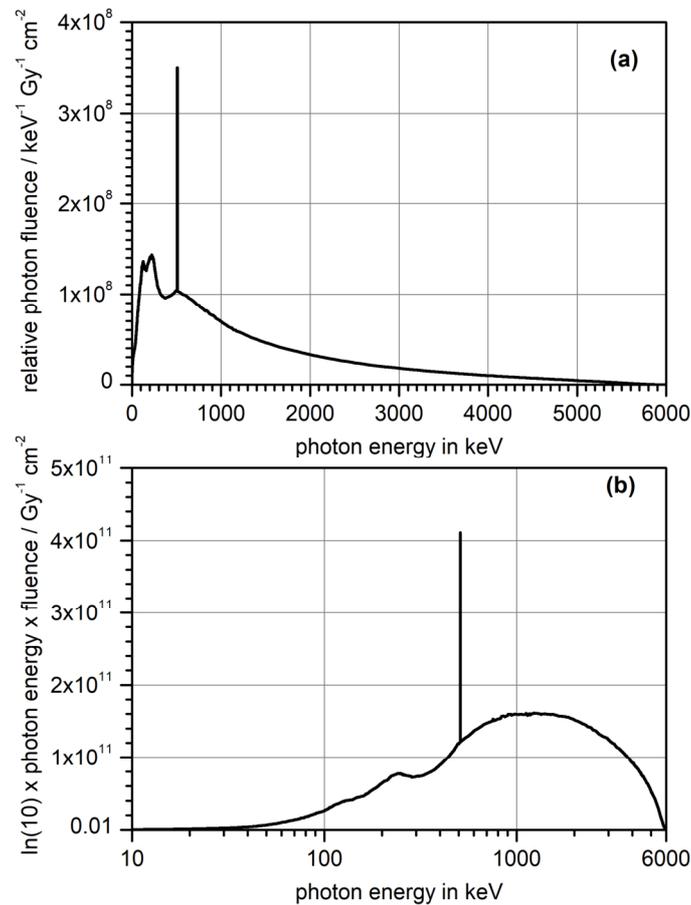

Fig. 3: (a) Photon fluence spectrum in 5 cm of water for a 6 MV linac source. The data were taken from McMahon et al. (2011) and corrected (see text). (b) The same photon fluence spectrum plotted in "microdosimetry style" with a logarithmic $x$-axis and a $y$-axis showing the fluence multiplied by the photon energy and the natural logarithm of 10. In this way, the area under the curve is representative of the contribution to the total photon fluence of the different energy regions.

The photon fluence spectrum corrected for this error was used in the further analysis and is shown in Fig. 3. That this photon fluence spectrum has the correct order of magnitude can be seen by the following consideration: The mean energy of the photon spectrum is about 1.3 MeV. The corresponding mass energy transfer coefficient is about 0.03 cm²/g (Hubbell and Seltzer 2004). This gives a dose-to-fluence ratio under secondary particle equilibrium of about $6\times10^{-12}$ Gy cm². Therefore, a photon fluence in the order of $2\times10^{11}$ cm$^{-2}$ is needed to produce a dose of 1 Gy in water. The integral under the photon fluence curve in Fig. 3 is of this magnitude.

### *Estimate for the magnitude of the local dose around a GNP*

The local dose around a GNP was estimated based on the expected number of photon interactions in the GNP and on literature data for the energy deposition or local dose around a GNP. The first dataset was from a multi-center comparison of simulated dose enhancement around GNPs under X-ray irradiation (Rabus, Li, Villagrasa, et al. 2021). The second dataset was taken from Fig. 2 of the publication of McMahon et al. (2011), which shows results for a 6 MV linac irradiation on 2 nm GNPs in 5 cm depth of water.

For a uniform isotropic photon field of fluence $\Phi$ (particles per area), the expected number of photon interactions, $\bar{n}_p$, in a GNP is given by Rabus, Li, Nettelbeck, et al. (2021):

$$\bar{n}_p = \Phi \times \langle \mu_g \rangle_{\Phi_E} \times V_g \qquad (7)$$

where $V_g$ is the volume of the GNP and $\langle \ \rangle_{\Phi_E}$ indicates a weighted average with respect to the spectral distribution of the photon fluence, $\Phi_E$. Therefore, for a uniform isotropic photon field, the probability of a photon interaction in the GNP does not depend on the shape of the GNP volume.

For an isotropic point source of a given radiant intensity $dN/d\Omega$ (particles per solid angle), the fluence of emitted particles is inversely proportional to the square of the radial distance. For the two GNP shapes of sphere and cube with point sources located at 1 nm from the GNP surface, there is therefore a different photon fluence at the GNP center. This fluence value is in the order of 50 % larger for the cube than for the sphere.

However, since the source is that close to the GNP, the value at the center is not representative for the whole GNP. Instead, the interaction probability for a point source depends on the mean chord length, that is, the expectation of the chord length (CL) distribution. The CL is the length of the path of a beam inside a given volume. CL distribution and the mean chord length depend on the GNP shape. This is also the case for the uniform isotropic irradiation geometry, where the mean CL can be obtained according to Cauchy's theorem (Kellerer 1971) as $4/3 \times r$ and $2/3 \times s$ for the sphere and the cube, that is, 20 nm and about 16.1 nm, respectively. However, owing to the uniform fluence distribution, the different chords cover the volume uniformly. In contrast, for a point source outside the volume, the expected number $\bar{n}_p$ of photon interactions in the two GNP shapes is given by Eq. (8).

$$\bar{n}_p = \frac{dN}{d\Omega} \times \langle \mu_g \rangle_{\Phi_E} \times \langle \ell \rangle_\Omega \qquad (8)$$

In Eq. (8), $\mu_g$ is the linear attenuation coefficient of gold, $\ell$ is the length of the chord inside the GNP of a beam starting at the point source, and $\langle \ \rangle_{\Phi_E}$ and $\langle \ \rangle_\Omega$ indicate the average over the photon energy spectrum and the full solid angle, respectively.

### *Chord length distributions*

In the work of Gray et al. (2023), simulations of electrons emitted from the GNP were conducted with the photon fluence obtained in the first simulation and assuming six isotropic point sources slightly outside the GNP. To obtain the mean CL for a sphere of radius $r$ = 15 nm and a cube of the same volume (that is, a side length $s$ of about 24.2 nm), the CL distributions were determined by random-sampling $10^8$ radial beams from a point. The point was located outside the GNP at a distance $d$ = 1 nm from the GNP surface along one of the symmetry axes of the geometrical shape. (Considering only one point is sufficient due to the symmetry of the geometry.)

The cosine of the polar angle (with respect to the vector from the source point to the center of the GNP) was uniformly sampled in the interval between $1 - \cos\theta_m$ and 1. $\theta_m$ is the maximum polar angle for which a beam intersects the GNP. (For the sphere, $\theta_{m,s}$, and for the cube, $\theta_{m,c}$ from Eq. (4) were used, respectively.) Exploiting the symmetry, the azimuthal angle was uniformly sampled between 0 and $\pi/4$ for the cube and was ignored for the sphere. The resulting CL distributions were normalized by multiplying with $(1 - \cos\theta_m)/(2n\Delta L)$, where $n$ is the number of beams and $\Delta L$ is the bin size of the CL histogram.

CL distributions were also determined for the case of uniform isotropic irradiation. For the cube, $10^8$ chord lengths were obtained by first random-sampling a beam direction with the cosine of the polar angle uniformly distributed between -1 and 1 and the azimuth uniformly distributed between 0 and $2\pi$. Then a random point was sampled in a plane perpendicular to this direction within a circle around the center of the cube of radius equal to $s \times \sqrt{3}/2$ ($s$ is the side of the cube). To correct for the fraction of beams not intersecting the cube, the chord length distribution was normalized by multiplying with $1/(n_{>0}\Delta L)$, where $n_{>0}$ is the number of beams intersecting the cube and $\Delta L$ is the bin size of the CL histogram. For the sphere, the CL histogram was directly constructed from the known analytical expression (Kellerer 1971) such that the frequency density for the $k^{th}$ bin was calculated as $(2k + 1) \times \Delta L/(2r)^2$.

From the CL distributions, the mean CLs were determined by multiplying the CL value by the frequency density and the CL bin width and summing over all bins. The conditional mean CL for beams intersecting the target was also determined by dividing the mean CL by the sum of the frequencies. For the cube and the point source, the actual solid angle was determined from the sum of the frequencies for non-zero CL.

### *Uncertainty of the radiation enhancement factors*

In the study of Gray et al. (2023), the parameters of the L-Q model of cell survival were obtained by fitting the model to the observed survival rates. Since measurements were only performed at two dose values, the parameters of the L-Q model can be directly calculated by solving the set of two linear equations given by Eq. (9).

$$-\ln S_{x_g,3Gy} = \alpha_{x_g} \times 3\ Gy + \beta_{x_g} \times 9\ \text{Gy}^2$$
$$-\ln S_{x_g,6Gy} = \alpha_{x_g} \times 6\ Gy + \beta_{x_g} \times 36\ \text{Gy}^2 \quad . \tag{9}$$

In Eq. (9), ln denotes the natural logarithm, and $\alpha_x$ and $\beta_x$ are the parameters of the L-Q model curve of cell survival. $S_{x_g,3Gy}$ and $S_{x_g,6Gy}$ are the observed survival rates for doses of 3 Gy and 6 Gy, respectively, at a mass fraction $x_g$ of gold. The dose $D_{x_g}(S)$ that produces a given survival level $S$ is then obtained as

$$D_{x_g}(S) = -\frac{\alpha_{x_g}}{2\beta_{x_g}} + \sqrt{\left(\frac{\alpha_{x_g}}{2\beta_{x_g}}\right)^2 - \frac{\ln S}{\beta_{x_g}}} \quad . \tag{10}$$

The values of $S_{x_g,3Gy}$ and $S_{x_g,6Gy}$ and their uncertainties were read from Fig. 7 of Gray et al. (2023) using the inkscape tool (https://inkscape.org). Using Eqs. (1) and (10) with the solution of Eq. (9), the REFs and their uncertainties were calculated.

In addition, a simultaneous non-linear regression of all data with the model function shown in Eq. (11) was also performed.

$$-\ln S_{x_g,D} = \alpha \times (1 - x_g + \gamma x_g)D + \beta \times \left[(1 - x_g + \gamma x_g)D\right]^2 \quad . \tag{11}$$

This model function assumes that the different survival curves are only due to a dose enhancement, expressed by the additional parameter $\gamma$ (Eq. (6)), while the other two model parameters are independent of $x_g$. For this approach the uncertainties of the model parameters were also determined.

**Results**

*Upper bound for the DEF in the simulations*

Using the spectrum shown in Fig. 3(a) for the photon fluence reported in McMahon et al. (2011) for 1 Gy cm², dose-to-fluence ratios of 1.09 Gy cm² and 2.23 Gy cm² are obtained for water and gold, respectively. The deviation of the first value from the expected value of 1 Gy cm² is about 9% and indicates the overall uncertainty of the procedure used here. This uncertainty includes the limited accuracy of extracting the data by digitization from a printed figure as well as interpolation of the extracted data points and of the tabulated mass energy transfer coefficients in Hubbell and Seltzer (2004).

Using the values given above in Eq. (6) gives a value of $\gamma_{g,w}$ of approximately 2, so that from Eq. (5), a maximum DEF of 1.00055 is expected for the 6 MV photon spectrum at a mass fraction of gold of $2.7 \times 10^{-4}$, that is, for a single spherical GNP of 15 nm radius in a 1 μm³ volume of water. For the 0.1 % and 0.15 % mass fractions used in the experiments of Gray et al. (2023), the estimated maximum DEFs are about 1.002 and 1.003, respectively. The deviation of these DEF values from unity has a relative uncertainty in the order of 10%.

Since the photon energy spectrum shown in Fig. 3 is dominated by high-energy photons, the correction to be applied to account for energy absorption in the GNP can be estimated from the results shown in Koger and Kirkby (2016) to be maximum 5 %. Therefore, the aforementioned DEFs are representing the expected order of magnitude for the 6 MV spectrum.

*CL distributions and mean CLs*

Fig. 4 shows a comparison of the CL distributions for uniform isotropic irradiation and irradiation of a GNP from the point source considered in the second simulation steps of Gray et al. (2023). In both cases a steep increase can be seen at about 24 nm for the CL distribution of the cube. This is not an artefact but rather reflects the fact that starting with this CL, lines leaving the cube through the back side (as seen from the source point) contribute to the distribution in addition to lines leaving the cube on the side.

For both GNP shapes significant changes can be seen between the two irradiation geometries. For the sphere the linear distribution for uniform isotropic irradiation turns into a curve with saturation behavior for the point source. For the cube a peak appears at about half the side, and the range of CL values is reduced for the point source. This is easily understandable since some beam geometries such as beams passing through opposing corners or edges of the cube are not possible for the point source geometry.

The frequency density of the CL distribution relating to the point source is lower than that of the uniform isotropic case. This reflects the fact that for the point source only a fraction of the full solid angle is covered by beams intersecting the GNP (Fig. 1).

The mean CLs for the uniform isotropic irradiation of the sphere and the cube are obtained as 20 nm and about 16.1 nm, respectively, in accordance with Cauchy's theorem. For the point source, the mean CLs amount to about 7.08 nm for the cube and 5.84 nm for the sphere (Table 1). The ratio of these mean CLs for the cube and the sphere is approximately 1.21. This means that the mean CL and the interaction probability of a photon is about 21 % higher for the cube than for the sphere. Thus, a major part of the larger dose enhancement from the cube of about 30% (DEF of 8 vs. DEF of 6 for the sphere) appears to be due to a bias introduced by the way the photon fluence was sampled in the work of Gray et al. (2023).

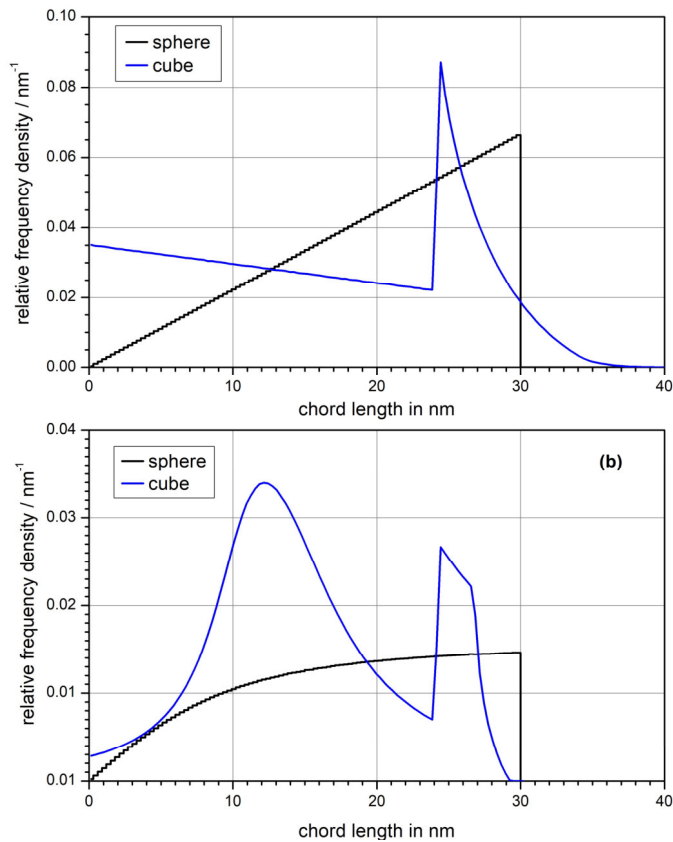

Fig. 4: Chord length distribution within a sphere of 30 nm diameter and a cube of the same volume for (a) a uniform isotropic radiation field and (b) an isotropic point source located outside the respective volume at a radial distance of 1 nm from the sphere surface and a linear distance along the surface normal from the center of one of the squares forming the surface of the cube.

Table 1: Mean chord lengths for the cubic and spherical GNP in the point source geometry and the estimated resulting probabilities of photon interaction in the GNP for the photon fluence spectrum shown in Fig. 3. The uncertainties of the mean CLs and their ratio are the standard deviation from 100 batches of $10^6$ random samples. The uncertainty of the interaction probability is dominated by the uncertainty of the fluence-mean of the linear attenuation coefficient $\langle \mu_g \rangle_{\Phi_E}$.

| GNP shape | sphere | cube |
|---|---|---|
| Mean CL / nm | 5.837 ± 0.003 | 7.076 ± 0.004 |
| Interaction probability | (1.50 ± 0.15) × 10$^{-5}$ | (1.82 ± 0.18) × 10$^{-5}$ |
| Ratio mean CL cube to sphere | 1.212 ± 0.004 ||

*Local dose per photon*

For the case of uniform isotropic irradiation, the probability of a photon interaction in the GNP calculated with Eq. (7) is 6.36×10$^{-5}$ for a fluence producing a dose of 1 Gy (about 1.9×10$^{11}$ cm$^{-2}$). For the point source geometry used in the second simulation of Gray et al. (2023), the probability of a photon interaction occurring in the GNP as obtained from Eq. (8) is 1.50×10$^{-5}$ per emitted photon for the sphere. For the cube, the corresponding probability is about 1.82×10$^{-5}$ (Table 1).

In the work of Rabus, Li, Villagrasa, et al. (2021), the mean energy imparted by electrons emitted from the GNP under X-ray irradiation was found to be more or less constant at distances from the GNP surface between about 150 nm and 1000 nm. Since the GNPs considered in that work had diameters of 50 nm and 100 nm, this translates into distances from the GNP center of 200 nm or more. The amount of energy imparted per 10 nm spherical shell was in the order of 40 eV per photon interaction in the GNP. A 10 nm thick spherical shell of 200 nm mean radius has a volume of about 5×10$^{-3}$ μm$^3$, which corresponds to a mass of about 5×10$^{-18}$ kg. Therefore, the estimated dose per photon interaction in the GNP at 200 nm from the GNP is about 1.3 Gy. For the GNP sizes considered by Gray et al. (2023), the estimated dose at this distance amounts to about 2.5×10$^{-5}$ Gy.

Mc Mahon et al. (2011) presented in their Fig. 2 the dose as a function of radial distance from the GNP center. At 200 nm distance a dose of 0.18 Gy per ionization in the GNP was found for the 6 MV linac spectrum and a GNP of 2 nm diameter. Applying this value of dose per interaction to a spherical GNP of 30 nm diameter gives an estimated dose at 200 nm from the GNP center of about 3.5×10$^{-6}$ Gy.

That the two estimates of the local dose are different is easily understood. The data from the work of Rabus, Li, Villagrasa, et al. (2021) were determined for irradiation of the GNP with 50 kVp and 100 kVp X-ray spectra. The photons of these spectra are in an energy range in which photoabsorption is the dominant process of photon interaction in gold (Rabus et al. 2019). The spectra have a large component of low-energy photons that produce photoelectrons of energies comparable to the energies of the gold L-shell Auger electrons. In contrast, the photon fluence spectrum used by McMahon et al. (2011) (shown in Fig. 3) has only a small contribution of photons with energies below 100 keV. This means that most photoelectrons have energies much higher than those of the gold L-shell Auger electrons and, thus, a much smaller energy loss in the vicinity of the GNP. Furthermore, about 50% of the photons have energies in the range above 500 keV, where incoherent Compton scattering is the dominant interaction process in gold (Rabus et al. 2019). Therefore, one may expect a significant reduction of the dose around the GNP when moving from an X-ray photon spectrum, as used by Rabus, Li, Villagrasa, et al. (2021), to a 6 MV photon spectrum as used by McMahon et al. (2011).

The value shown in Fig. 5 of Gray et al. (2023) for the 6 MV irradiation and the spherical GNP is about $1.8 \times 10^{-25}$ Gy. Assuming that the estimate derived from the data of McMahon et al. (2011) is the more relevant one, it appears that these values are too small by a factor in the order of $5 \times 10^{-20}$.

*L-Q model parameters and REFs*

The results of the analysis of the data for cell survival presented in Fig. 7 of Gray et al. (2023) are listed in Table 2. The first row shows the gold concentration (mass fraction). The second and third rows list the parameters of the L-Q model applied to the data of each mass fraction. The values of $\alpha_{x_g}$ show an increasing trend with increasing $x_g$, whereas the values of $\beta_{x_g}$ decrease.

The next two rows give the intermediate results for the doses needed to produce a cell survival rate of 0.3 and 0.6, respectively. These values show a decreasing trend with increasing $x_g$. The ensuing two rows are the corresponding REF values. Similar to the values reported by Gray et al. (2023), the REF values are higher for the higher survival rate, and the deviation from unity doubles for 0.15 % mass fraction compared to 0.1 %. However, the uncertainties are so large for all parameters that the changes are not statistically significant.

The last three rows of Table 2 show the parameters obtained by fitting Eq. (11) to all data. The values of $\alpha$ and $\beta$ are approximately equal to the means of $\alpha_{x_g}$ and $\beta_{x_g}$, respectively. This is expected, as is the observation that the uncertainties are much smaller than for the individual $\alpha_{x_g}$ and $\beta_{x_g}$. The value of the parameter $\gamma$ is comparatively large, given that for the 6 MV photon spectrum a factor of approximately 2 was found for the constant of proportionality between (*DEF*-1) and $x_g$ (Eq. (5)). For the 18 MV radiation, the photon fluence spectrum is expected to have more high-energy contributions for which the mass energy absorption coefficient should be smaller than for the fluence spectrum deriving from 6 MV linac radiation.

**Discussion**

*Magnitude of the simulated local dose and DEF*

The contradiction between the radiobiological experiments and the simulations in Gray et al. (2023) seems to be due to incorrect results obtained from the simulations. The reported DEF values in the 1 μm³ water volume are much higher than what one expects from the energy transfer coefficients of photons. For the 6 MV photon fluence spectrum from the work of McMahon et al. (2011), a maximum DEF was estimated to be about 1.00055 for a mass fraction of gold of 2.7×10$^{-4}$ (1 μm³ water volume containing a cubic GNP of 24 nm side). Similarly, DEFs of about 1.002 and 1.003 would apply for the cell experiments with 0.1 % and 0.15 % mass fraction of gold, respectively, if they were conducted with 6 MV irradiation instead of 18 MV.

Table 2: Results of the analysis of the cell survival data presented in Gray et al. (2023) following the approach of determining the radiosensitivity enhancement factor (REF) according to Eq. (1): $\alpha$ and $\beta$ are the parameters of the L-Q model, $D_{x_g}$ is the dose corresponding to a given survival level according to Eq. (7). The last three rows are the best fit parameters of all data to Eq. (11).

| $x_g$ | 0 % | 0.1 % | 0.15 % |
|---|---|---|---|
| $\alpha_{x_g}$ / Gy$^{-1}$ | 0.16 ± 0.06 | 0.20 ± 0.07 | 0.24 ± 0.07 |
| $\beta_{x_g}$ / Gy$^{-2}$ | 0.070 ± 0.006 | 0.063 ± 0.008 | 0.060 ± 0.008 |
| $D_{x_g}(0.3)$ / Gy | 3.18 ± 0.20 | 3.04 ± 0.23 | 2.90 ± 0.23 |
| $D_{x_g}(0.6)$ / Gy | 1.81 ± 0.20 | 1.66 ± 0.22 | 1.54 ± 0.21 |
| $REF_{x_g}(0.3)$ | - | 1.05 ± 0.10 | 1.10 ± 0.11 |
| $REF_{x_g}(0.6)$ | - | 1.09 ± 0.19 | 1.18 ± 0.21 |
| $\alpha$ / Gy$^{-1}$ | 0.20 ± 0.02 | | |
| $\beta$ / Gy$^{-2}$ | 0.062 ± 0.04 | | |
| $\gamma$ | 20.6 ± 8.4 | | |

In the work of Gray et al. (2023), the DEF is stated to be determined from the microscopic simulations with GNP and the dose obtained from the first macroscopic simulation for a volume of water without GNP inside. The latter dose value was obtained under conditions of secondary electron equilibrium (Gray et al. 2021). However, for the dose with the GNP, only the energy imparted by electrons emitted from the GNP (as determined in the second simulation) was scored. This simulation only considered the single 1 μm³ volume containing a GNP and, therefore, a situation of charged particle disequilibrium. This is expected to result in a large underestimation of the dose in the 1 μm³ volume when GNPs are present. Nevertheless, the reported DEFs were much larger than unity, which is generally an indication that both dose without and with GNP were determined under secondary particle disequilibrium.

The DEFs of the average dose in the 1 μm³ water volume deviated from unity by more than a factor of 10,000 with respect to the deviation from unity of the above upper bound for the 6 MV spectrum. In contrast, the values for the local dose per photon emitted from the point source shown in Fig. 5 of Gray et al. (2023) are about 20 orders of magnitude smaller than the values estimated based on the two different approaches used. Since not much detail is given in the work of Gray et al. (2023) on how the different steps of the simulations were linked, the reason for these obviously wrong results remains obscure.

One possible source of error is the conversion of the photon fluence (particles per area) from the first simulation into the radiant intensity (particles per solid angle) of the point sources used in the second simulation. The most probable explanation for the conflicting directions of the deviation from the expected values or orders of magnitude for the dose in a 1 μm³ water volume containing a GNP and for the local dose around this GNP is the occurrence of two errors related to normalization.

*Dependence of the dose enhancement on nanoparticle shape*

The analysis of the chord length distributions and mean chord lengths for the point source considered by Gray et al. (2023) showed that this irradiation geometry produces a higher probability of a photon interaction in the cube-shaped GNP. The effect was an about 21% increase for the cubic GNP compared to the spherical one. This accounts for a large proportion of the 30 % higher DEF of the cubic compared to the spherical nanoparticle. The unresolved issues regarding the magnitude of local absorbed dose and average dose in the 1 μm³ water volume leads to the question whether the differences between the cube and the sphere still remain when these issues are resolved.

The photon field can be expected to be uniform over the small dimensions of the nanoparticle. Therefore, the probability of photons interacting in the nanoparticles is expected to be proportional to the nanoparticle's volume (Rabus et al. 2019) and, hence, should be the same for the two shapes of nanoparticle. (At least when either the photon fluence is isotropic or the non-spherical nanoparticles are randomly oriented.) Thus, an increased dose for the cube as compared to the sphere should be solely due to the difference in the emitted electron spectra.

The surface-to-volume ratio is about 25% higher for the cubic nanoparticle compared with the spherical one. This will lead to an enhanced emission of low-energy electrons (Au M- and N-shell Auger electrons and low energy secondaries), which are stopped within the first 150 nm from the GNP surface (Rabus, Li, Villagrasa, et al. 2021). This contribution can be expected to show an increase proportional to the increased surface area. In contrast, no significant change with GNP shape is expected for the energy imparted by all higher-energy electrons in the 1 μm³ water volume around the GNP.

This is because the energy imparted per unit path by electrons within higher distances up to 1 μm from the GNP (mainly from gold L-shell Auger electrons and electrons of similar energies) was found to be almost constant and to be approximately independent of the photon energy spectrum for X-ray spectra. For higher photon energy spectra, Compton and photoelectrons produced in the GNP will generally have so high energies that their energy loss within the first 1 μm around the GNP is negligible, such that the energy imparted is only from gold L-shell Auger electrons in this case.

The energy imparted in the 1 μm³ volume is composed of the contribution of low-energy electrons from the GNP, which is expected to increase with surface area, and the contribution of electrons of higher energies, which is independent of the GNP shape. Therefore, the dose in 1 μm³ water volume around a cubic GNP is expected to be increased compared to the case of a spherical GNP. Consider a sphere of 1 μm³ volume with a radius of 625 nm. From the data shown in Fig. 6 of Rabus, Li, Villagrasa, et al.

(2021), approximately one third of the energy imparted within the first 625 nm around the GNP is deposited within the first 100 nm. Two thirds of this energy can be attributed to low-energy electrons that are stopping in the range, which makes up about 25 % of the total energy imparted in a sphere with 625 nm radius. When this contribution increases by 25 %, the overall increase of dose in the volume is in the order of 6 %. This is of similar magnitude as the difference between the 30 % increase in DEF reported by Gray et al. (2023) and the 21 % bias introduced by the irradiation geometry in their second simulation step. This means that there may remain a small benefit of using cubic instead of spherical GNPs.

*Significance of the REF values*

The analysis presented in the subsection "L-Q model parameters and REFs" showed that the uncertainties associated with the radiobiological data are too high to make a statistically significant case. The alternative approach of simultaneously analyzing all data with the assumption that the changes seen can be described only by dose enhancement gave a proportionality factor for the excess dose contribution from the GNPs that is an order of magnitude higher than the value estimated for the 6 MV irradiation. This is implausible, since for the higher photon energy spectrum, incoherent scattering is more important and produces higher-energy electrons than Auger electrons emitted after photoabsorption. Therefore, if the reduced survival was only due to an increase of the average dose, the parameter $\gamma$ is expected to be smaller for the 18 MV irradiation. This expectation is also supported by the simulations of Gray et al. (2023), where a higher effect was found for the 6 MV spectrum. While the REF values and their changes are not statistically significant, the alternative analysis presented here shows that the reduced survival in the presence of GNPs is not an effect of the average dose enhancement and that other factors, such as the local dose enhancement (McMahon et al. 2011; Butterworth et al. 2012; Zygmanski and Sajo 2016; Kirkby et al. 2017; Hullo et al. 2021) or chemical effects have to be considered as well.

**Conclusions**

The work of Gray et al. (2023) contains a contradiction between the results of their radiobiological assays and Monte Carlo simulations as well as between different results obtained from these simulations. As discussed here, the reported dose enhancement factors for one GNP in a 1 µm³ volume are inconsistent with upper limits estimated from the principle of energy conservation. The deviation of the DEFs from unity (no dose enhancement) are higher by a factor of about 10,000 than the difference between the upper bounds and unity. At the same time, the data presented by Gray et al. (2023) for the local dose around a GNP are about 20 orders of magnitude too low. This suggests that these results are compromised by two different normalization errors.

    The analysis presented here further indicates that the large variation of dose enhancement with nanoparticle shape (for the same volume) seems to be mostly due to a bias introduced by the simulation setup for the microscopic simulation of photons interacting with the GNPs. In addition, the reanalysis of the cell survival data revealed that the uncertainties of the radiosensitization enhancement factors obtained by fitting the L-Q model to just two data points are so large that the differences appear not to be statistically significant.

## Acknowledgements

L.T. acknowledges funding by (source to be inserted after peer review)